\documentclass[letterpaper,prl,twocolumn,showpacs,superscriptaddress]{revtex4}

\usepackage{graphicx}
\usepackage{amsmath}

\newcommand{\eg} {{e.g., }}

\newcommand{\ie} {{i.e., }}
\newcommand{\Op} {\hat{O}}
\newcommand{\pd} {\partial}
\newcommand{\vecj} {{\bf j}}

\begin{document}

\title{Localized Rayleigh Instability in Evaporation Fronts}

\author{Haim Diamant}
\affiliation{School of Chemistry, 
Sackler Faculty of Exact Sciences, Tel Aviv University,
Tel Aviv 69978, Israel}
\affiliation{The Racah Institute of Physics, The Hebrew University,
Jerusalem 91904, Israel}

\author{Oded Agam}
\email{agam@phys.huji.ac.il}
\affiliation{The Racah Institute of Physics, The Hebrew University,
Jerusalem 91904, Israel}

\date{\today}

\begin{abstract}
  A qualitatively different manifestation of the Rayleigh instability
  is demonstrated, where, instead of the usual extended undulations
  and breakup of the liquid into many droplets, the instability is
  localized, leading to an isolated narrowing of the liquid
  filament. The localized instability, caused by a nonuniform
  curvature of the liquid domain, plays a key role in the evaporation
  of thin liquid films off solid surfaces.
\end{abstract}


\maketitle

The Rayleigh instability of slender liquid bodies, driven by surface
tension, is part of everyday experience and has been systematically
studied for well over a century \cite{Eggers97,deGennesBook}. Its
manifestations are diverse, ranging from the breakup into droplets of
inviscid, viscous, and viscoelastic liquid jets and bridges
\cite{Eggers97} to the pearling of fluid membranes \cite{BarZiv}.  The
liquid bodies considered in earlier studies were usually
translation-invariant along their long dimension, resulting in
unstable modes which were extended \cite{Eggers97,deGennesBook} or had
a steadily propagating front \cite{Powers}. The propagation of forced
perturbations from a fixed nozzle were studied as well
\cite{Eggers97}. In this Letter we investigate a qualitatively
different scenario of the Rayleigh instability, in which the
translation invariance is broken by a nonuniform curvature, and show
that the fastest-growing mode of this instability is localized. The
localization is reminiscent of the problem of a quantum particle
moving inside a curved stripe \cite{daCosta,Exner}.  When the latter
system is transformed into an effective one-dimensional problem, an
attractive potential emerges, whose minimum is located at the point of
maximum curvature, giving rise to bound (localized) states.

\begin{figure}[b]
\centerline{\resizebox{0.4\textwidth}{!}
{\includegraphics{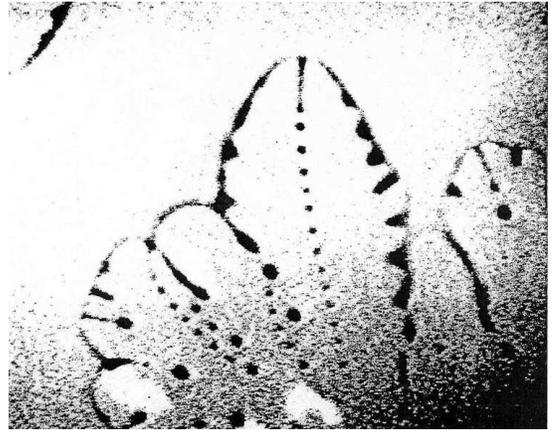}}}
\caption[]
{Doublon patterns observed in the evaporation of water off a clean
  mica surface \cite{ft_Lipson}.  Fingers of a molecularly thin water
  film grow into a thicker film. A local indentation at the finger tip
  leads to splitting and the formation of a liquid spine that
  subsequently breaks into droplets. The fingers are hundreds $\mu$m
  wide.}
\label{fig_doublon}
\end{figure}

Thin liquid bodies are abundant in phenomena related to wetting of
solid substrates \cite{deGennesBook,EggersRMP}.  Two processes are
particularly relevant to the current work: (i) the dewetting of a
nonvolatile, nonwetting film \cite{deGennesBook,Brochard91,Seeman01};
and (ii) the evaporation of a volatile, totally wetting film
\cite{Lipson94,Lipson98,Lipson03,Lipson04}. Both processes exhibit the
kinetics of a first-order transition, where dry domains [in (i)], or
domains covered by a molecularly thin liquid [in (ii)], nucleate and
grow into a much thicker film. Importantly, in both processes
the dewetting front has a long, slender rim of excess fluid
\cite{deGennesBook}.  The growing domains may have a stable circular
boundary \cite{Brochard91,Lipson94} or evolve through elaborate
instabilities and patterns
\cite{Lipson94,Lipson98,Lipson03,Lipson04,Gu02}. An analogy has
recently been drawn between the pattern formation in the volatile case
and Saffman-Taylor viscous fingering in a Hele-Shaw cell having a
time-varying thickness \cite{Oded}. The additional dynamics of the
liquid rim at the domain boundary, however, crucially affects the
selection of patterns in experiments, \eg the doublon pattern shown in
Fig.\ \ref{fig_doublon}, which is uncommon in Saffman-Taylor fingering
\cite{Lipson98,Oded}.

\begin{figure}[tbh]
\centerline{\resizebox{0.4\textwidth}{!}
{\includegraphics{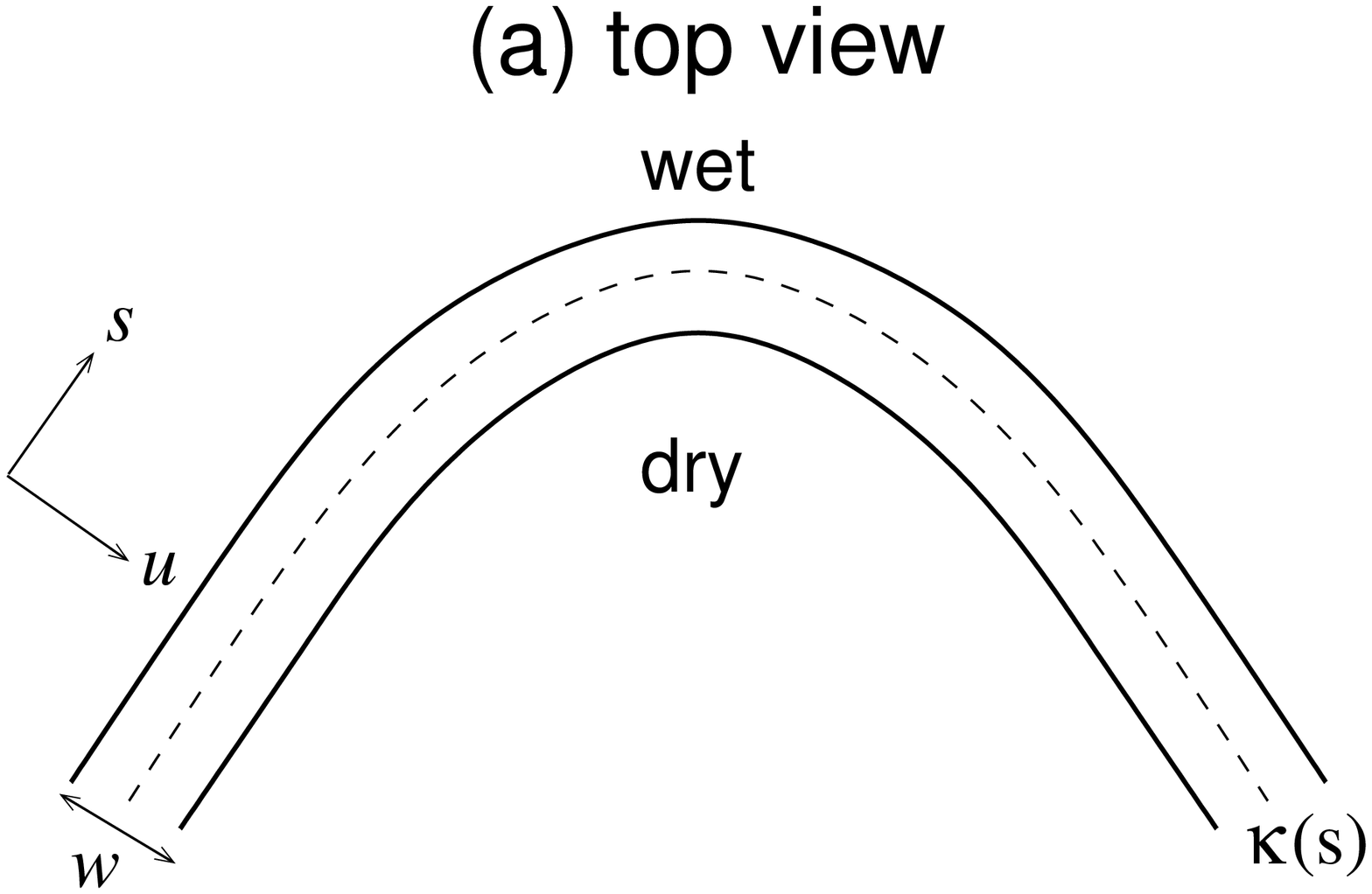}}}
\vspace{0.15cm}
\centerline{\resizebox{0.4\textwidth}{!}
{\includegraphics{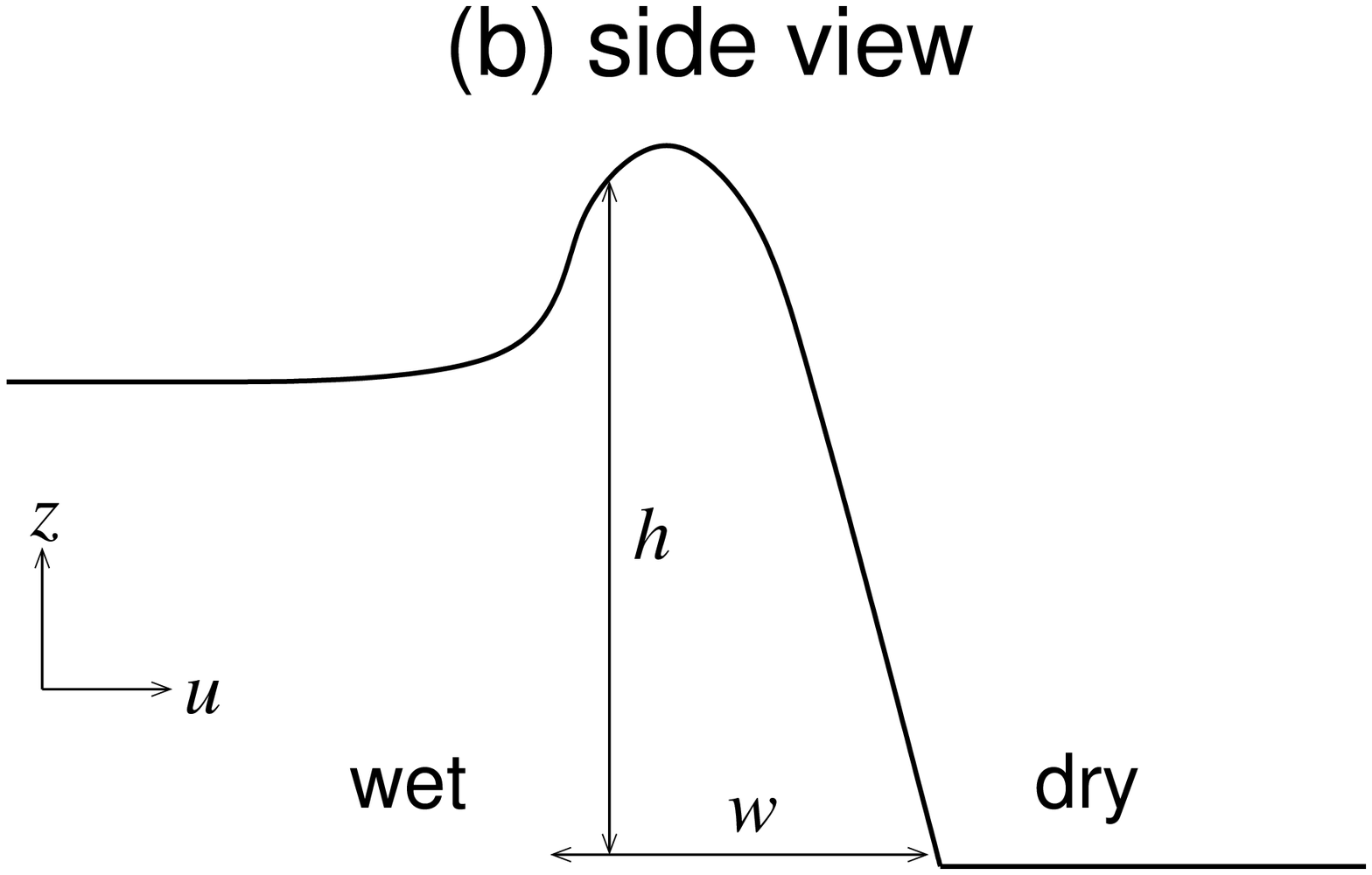}}}
\caption[]
{Schematic view of the system and its parametrization. A curved
liquid rim of uniform width $w$, nonuniform curvature $\kappa(s)$, and
height profile $z=h(s,u)$, lies at the interface between a thick liquid
film (wet domain) and a much thinner one (dry domain).}
\label{fig_scheme}
\end{figure}

We model the droplet's rim as a long, curved liquid strip of uniform
width $w$.  It is parametrized using the locally orthogonal triad
$(s,u,z)$ as depicted in Fig.\ \ref{fig_scheme}, where
$u\in[-w/2,w/2]$ and $z\in[0,h]$, $h(s,u)$ being the local liquid
height. The rim is in contact with the wet and dry domains at $u=-w/2$
and $w/2$, respectively. The local curvature of the centerline
$(s,0,0)$ is denoted by $\kappa(s)$ and taken as positive when the rim
curves away from the dry domain. (In Fig.\ \ref{fig_scheme}
$\kappa<0$.)  We assume separation of time scales between the growth
of the finger and the faster development of the rim instability, as
observed in the evaporation experiments. Thus, although the rim itself
is a dynamic effect, $w$ and $\kappa(s)$ can be taken as
time-independent.

Within the lubrication approximation ($|\nabla h|\ll 1$) \cite{Oron},
the equation of motion for the liquid is
\begin{equation}
  \pd_t h = -\nabla\cdot\vecj,\ \
  \vecj = -\frac{h^3}{3\eta} \nabla p,
\label{motion1}
\end{equation}
where $p$ is the liquid pressure, $\eta$ its viscosity, and spatial
derivatives are in the $(s,u)$ plane. In $p$ we include contributions
from the Laplace pressure and disjoining pressure $\Pi(h)$ (\ie
surface interactions) \cite{EggersRMP},
\begin{equation}
  p = -\gamma\nabla^2h - \Pi(h),
\label{pressure}
\end{equation}
where $\gamma$ is the surface tension of the liquid. Substituting in
Eqs.\ (\ref{motion1}) and (\ref{pressure}) a small perturbation of the
steady profile, $h=h_0+\psi$, linearizing in $\psi$, and neglecting
spatial derivatives of $h_0$, we obtain
\begin{equation}
  \pd_t\psi = \Op\psi,\ \
  \Op = -\Gamma_0q_0^{-4}\left( \nabla^4 + 2q_0^2 \nabla^2 \right).
\label{motion2}
\end{equation}
In  Eq.\ (\ref{motion2}) the Laplacian is given by \cite{Exner}
\begin{eqnarray}
  &&\nabla^2 = g^{-1/2}\pd_s g^{-1/2}\pd_s + g^{-1/2}\pd_u g^{1/2}\pd_u,
 \nonumber\\
  &&g(s,u) = [1+u\kappa(s)]^2,
\end{eqnarray}
and the following length and time scales appear:
\begin{equation}
  q_0^{-1} = [2\gamma/\Pi'(h_0)]^{1/2},\ \
  \Gamma_0^{-1} = 12\eta\gamma/(h_0^3 [\Pi'(h_0)]^2).
\label{scales}
\end{equation}

In the systems under consideration the disjoining pressure is usually
governed by van der Waals interactions \cite{EggersRMP}, whereby
$\Pi'=H/(2\pi h_0^4)$, $H\sim 10^{-13}$ erg being the Hamaker
constant. The length $b=(H/\gamma)^{1/2}$ is invariably of molecular
scale, $\sim 1$ nm. Hence, the value of $q_0^{-1}\sim h_0^2/b$ is
primarily determined by the rim thickness, $h_0$, which has very
different values in the dewetting and evaporation processes. In the
dewetting case of Ref.\ \cite{Brochard91} $h_0$ is about $50$
$\mu$m, leading to $q_0^{-1}$ of order meters and an unphysically long
$\Gamma_0^{-1}$. In the evaporation process of Ref.\
\cite{Lipson03}, by contrast, $h_0$ is of order $10$ nm, yielding
$q_0^{-1}$ of micron scale and $\Gamma_0^{-1}$ of order $10^{-2}$ s
(though the latter is highly sensitive to the thickness). The rim
width, $w$, is of mm scale in both cases; thus, $wq_0\ll 1$ for the
dewetting case, and $wq_0\gg 1$ for the evaporation one. As we shall
presently see, instability of the rim requires $wq_0\gtrsim 1$, which
clarifies the strikingly different dynamics observed in the two
processes.

A key feature of the system is that the rim separates domains of
differing properties (Fig.\ \ref{fig_scheme}). This implies asymmetric
boundary conditions at $u=\pm w/2$ and, consequently, sensitivity of
the results to the curvature direction (sign of $\kappa$).  At the
boundary with the thick liquid film we impose a fixed height and a vanishing
surface curvature,
\begin{equation}
  u=-w/2:\ \psi = 0,\ \ \nabla^2\psi = 0,
\label{bc1}
\end{equation}
thus ensuring that the pressure [Eq.\ (\ref{pressure})] changes
continuously between the rim and the film \cite{ft_mass}. At the
boundary with the dry domain we assume a fixed contact angle and a
vanishing outward current,
\begin{equation}
  u=w/2:\ \ \pd_u\psi = 0,\ \ \pd_u\nabla^2\psi = 0.
\label{bc2}
\end{equation}

The operator $\Op$, as defined by Eq.\ (\ref{motion2}) and the
boundary conditions (\ref{bc1}) and (\ref{bc2}), is hermitian.  In
general it does not commute with the Laplacian because of the
spatially varying $g(s,u)$. In cases where $\Op$ and $\nabla^2$ do
commute, the spectrum of $\Op$ (denoted by $\Gamma$) can be written in
terms of that of $\nabla^2$ ($\lambda$) as
$\Gamma=-\Gamma_0q_0^{-4}\lambda(\lambda+2q_0^2)$, and is thus bounded
from above by $\Gamma_0$. Hence, in such cases the fastest growing
mode has (at the most) a rate $\Gamma_0$ as given by Eq.\
(\ref{scales}).

In the simple case of a straight rim \cite{Lipson04}, $\kappa\equiv
0$, we have $[\Op,\nabla^2]=0$.  The eigenmodes of $\Op$ in this case
are extended,
\begin{eqnarray}
  \psi_{nq} &=& A e^{iqs + \Gamma_{nq}t} \sin[k_n(u+w/2)],
\label{mode1a}
\end{eqnarray}
where $A$ is an arbitrary amplitude, $q$ the wavenumber along the
rim, $k_n=\pi(n-1/2)/w$ ($n=1,2,\ldots$) the wavenumber in the
transverse direction, and
\begin{eqnarray}
  \Gamma_{nq} &=& \Gamma_0(q^2+k_n^2)[2q_0^2 -
  (q^2+k_n^2)]/q_0^4
\label{mode1b}
\end{eqnarray}
is the growth rate.
Unstable modes, having $\Gamma_{nq}>0$, are obtained for a
sufficiently wide rim, $wq_0>\pi/(2\sqrt{2})$; getting fastest growing
modes of finite wavelength requires the slightly stricter condition
$wq_0>\pi/2$. These modes have wavenumbers $q_n=(q_0^2-k_n^2)^{1/2}$
and the maximum growth rate $\Gamma=\Gamma_0$.
In the case of a circular rim of fixed curvature,
$\kappa\equiv\kappa_0\neq 0$, despite the nonuniform metric
$g=(1+\kappa_0u)^2$, $\Op$ and $\nabla^2$ still commute due to
rotational symmetry.  Hence, a finite uniform curvature cannot
accelerate the instability beyond the rate $\Gamma_0$.


We now address the interesting and practically relevant case of a
nonuniform curvature $\kappa(s)$.
Motivated by the evaporation experiments (Fig.\ \ref{fig_doublon} and
\cite{Lipson94,Lipson98,Lipson03,Lipson04})
and inspired by the quantum bound states (\ie localized eigenmodes of
$\nabla^2$) found in a similar curved geometry \cite{Exner}, we look
for a localized unstable mode of $\Op$, whose growth rate exceeds
$\Gamma_0$. To this end we employ a variational approach, which sets a
lower bound for the maximum rate according to
\begin{equation}
  \Gamma_{\rm max} \geq
  \bar{\Gamma}[\psi] \equiv \langle\psi|\Op|\psi\rangle /
  \langle\psi|\psi\rangle,
\label{Gammabar}
\end{equation}
$\psi(s,u)$ being any trial function that satisfies the boundary
conditions (\ref{bc1}) and (\ref{bc2}).

For a general form of $\kappa(s)$ it is difficult to construct a good
variational wavefunction which will satisfy the boundary
conditions. To simplify the analysis we assume that the curvature is
both small and slowly varying in space. In this regime the
wavefunction may be locally approximated by that of a circular rim
having curvature $\kappa(s)$. This observation leads to the following
choice of variational wavefunction:
\begin{eqnarray}
  \psi(s,u)&=&A\phi(s,u) e^{iq_1s} e^{-s^2/a^2},
\label{trial1}
\end{eqnarray}
where
\begin{eqnarray}
  \phi(s,u)&=&\{1 - \kappa(s)\frac{w}{\pi}
  [k_1(u+\frac{w}{2})-\frac{2}{\pi}]\} \sin[k_1(u+\frac{w}{2})]
 \nonumber\\
  &&- [2 \kappa(s) \frac{w}{\pi^2}k_1(u+\frac{w}{2})] \cos[k_1(u+\frac{w}{2})].
\label{trial2}
\end{eqnarray}
Here $k_1=\pi/(2w)$, $q_1=(q_0^2 -k_1^2)^{1/2}$, and the localization
length $a$ serves as a variational parameter \cite{ft_q1,ft_sin}.
Equation (\ref{trial2}) has been obtained from the asymptotic form of
the eigenmodes for a circular rim in the limit of small curvature,
while replacing $\kappa_0$ with the slowly varying curvature,
$\kappa(s)$.
For the sake of concreteness let us take
\begin{equation}
  \kappa(s) = \kappa_0 e^{-s^2/\sigma^2}.
\end{equation}
Our calculation is performed to leading order in two small parameters:
$\delta=\kappa_0w$ and $\epsilon=(q_1a)^{-1}$. (Note that the overall
turn of the rim, $\kappa_0\sigma$, may still be appreciable.) The
smallness of $\epsilon$, implying that the localized mode extends many
wavelengths away from the tip, is an ansatz to be confirmed below. In
addition, we assume $q_0w\gg 1$ (and thus $q_1w\gg 1$) to be safely
inside the unstable regime. The trial function of Eqs.\ (\ref{trial1})
and (\ref{trial2}) satisfies the boundary conditions for $\psi(-w/2)$
and $\pd_u\psi(w/2)$ exactly, while those for $\nabla^2\psi(-w/2)$
and $\pd_u\nabla^2\psi(w/2)$ are violated only at the orders
$\delta^2$ and $\delta\epsilon^2$, respectively.

Within this approximation we obtain from Eq.\ (\ref{Gammabar})
\begin{equation}
  \bar{\Gamma}(a) = \Gamma_0 \left[ 1 - 4\frac{q_1^2}{q_0^4}
  \left( \frac{1}{a^2} + \frac{\sqrt{2}\kappa_0\sigma}{wa} \right)\right].
\end{equation}
Maximizing $\bar{\Gamma}$ with respect to $a$ yields
\begin{eqnarray}
  a^* &=& -\sqrt{2} w / (\kappa_0\sigma) \nonumber\\
  \Gamma_{\rm max} &\geq& \bar{\Gamma}(a^*)
  = \Gamma_0 \left( 1 + \frac{2q_1^2\kappa_0^2\sigma^2}{q_0^4w^2} \right).
\label{main}
\end{eqnarray}
Thus, provided that the rim is negatively curved ($\kappa_0<0$ to get
$a^*>0$), the fastest-growing mode is localized and achieves a growth
rate larger than $\Gamma_0$.
The required curvature direction is in accord with the experiment,
where the instability occurs at the dry fingertips (\cite{Lipson98}
and Fig.\ \ref{fig_doublon}). Note that $\epsilon=(q_1 a)^{-1}\sim
|\kappa_0|\sigma/(q_1 w)\ll 1$, and $\Gamma_{\rm
  max}/\Gamma_0-1\sim\epsilon^2\ll 1$, which is consistent with the
aforementioned approximations. The main result, Eq.\ (\ref{main}),
seems robust to the choice of trial function \cite{ft_sin}.

\begin{figure}[tbh]
\vspace{0.6cm}
\centerline{\resizebox{0.45\textwidth}{!}
{\includegraphics{fig3a.eps}}}
\vspace{1.1cm}
\centerline{\resizebox{0.45\textwidth}{!}
{\includegraphics{fig3b.eps}}}
\vspace{0.4cm}
\centerline{\resizebox{0.45\textwidth}{!}
{\includegraphics{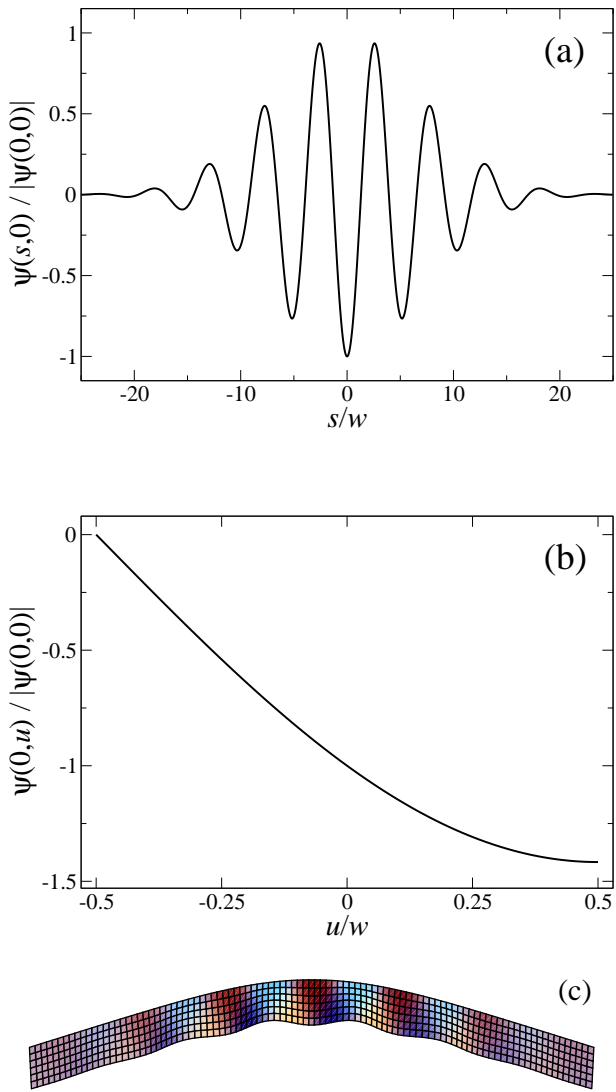}}}
\caption[]{Localized instability of a curved liquid rim. (a) Longitudinal profile
  of the height perturbation along the centerline, $\psi(s,0)$. (b)
  Transverse profile at the tip, $\psi(0,u)$. (c) (color online).
  Two-dimensional topography, $\psi(s,u)$. The parameters used are
  $\kappa_0/w=-0.02$, $\sigma/w=7$, and $q_0w=2$.}
\label{fig_mode}
\end{figure}

Two issues have remained unspecified in the discussion above. First,
for our choice of $\kappa(s)$ $\Op$ is symmetric in $\pm s$ and,
therefore, the actual unstable mode must have a definite parity [\ie
the $e^{iq_1s}$ factor in Eq.\ (\ref{trial1}) should be replaced by
either $\cos(q_1s)$ or $\sin(q_1s)$]. In the even case the
perturbation is maximum at the tip, whereas in the odd case it has a
node there. Within our approximation the difference in $\bar\Gamma$
between the two functions is minute. If the overall turn is relatively
small, $|\kappa_0\sigma|<\delta^{1/2}$, we have $a>\sigma$, and the
even perturbation is found to be slightly faster, with
$(\bar\Gamma_{\rm even}-\bar\Gamma_{\rm odd})/\Gamma_0 =
[8q_1^2\kappa_0^4\sigma^6/(q_0^4w^4)]e^{-q_1^2\sigma^2}$. An example
of an even unstable mode is shown in Fig.\ \ref{fig_mode}.
In the opposite case of $|\kappa_0\sigma|>\delta^{1/2}$ we have
$a<\sigma$, and the growth rate of the odd perturbation is slightly
higher, 
$(\bar\Gamma_{\rm odd}-\bar\Gamma_{\rm even})/\Gamma_0 = 2(q_1/q_0)^4
e^{-q_1^2 a^2/2}$.

The second question concerns the sign of the localized instability ---
whether it increases the liquid height at $s=0$ toward the formation
of a droplet, or decreases it toward pinch-off. A rigorous answer
requires nonlinear analysis that lies beyond the scope of the current
work. Yet, since the entire liquid film is unstable against
evaporation, it is plausible to expect that the rim should shrink at
its tip.
This is also the direction required to account for the experimentally
observed doublon patterns (Fig.\ \ref{fig_doublon}). By Darcy's law
the velocity of the interface increases with the local film height.
Hence, a local indentation at the tip of a dry finger advances more
slowly than its shoulders, leading to the shape of a finger split in
two by a narrow liquid spine.

This work has two direct experimental implications. The first relates
to the stable radial growth of domains, found in the dewetting of
nonvolatile liquid films \cite{deGennesBook,Brochard91,Seeman01}, vs.\
the unstable pattern formation observed in the evaporation of volatile
ones \cite{Lipson94,Lipson98,Lipson03,Lipson04}. We have shown that
the qualitatively different dynamics in these two types of experiment
can be related to the stability vs.\ instability of the accumulated
liquid rim at the domain boundary. The second implication concerns the
mechanism behind the patterns selected in the evaporation process. The
newly demonstrated effect --- a localized Rayleigh instability driven
by the surface tension of a nonuniformly curved liquid domain --- is
essential for the pattern formation as it suppresses the
Saffman-Taylor instability at the finger tips.
More broadly, since the relation between inhomogeneity and
localization is far more general (encountered, \eg in the effect of
defects on electron states in a solid), we expect related localization
effects to emerge in other scenarios of the Rayleigh instability.
Indicating these scenarios calls for further investigation. Such
localized instabilities, for example, may offer new possibilities to
control via curvature the precise location where a liquid filament is
to agglomerate into a drop or pinch off.

\begin{acknowledgments}
  We thank Yossi Avron, Eldad Bettelheim, Michael Elbaum, Steve
  Lipson, and Rachel Yerushalmi-Rozen for helpful discussions, and Tom
  Witten for valuable comments on the manuscript. HD wishes to thank
  the Racah Institute of Physics, Hebrew University, for its
  hospitality. This research has been supported in part by the Israel
  Science Foundation (ISF) under Grant Nos.\ 588/06 and 9/09.
\end{acknowledgments}




\begin{thebibliography}{99}
\bibitem{ft_mass}
The boundary condition (\ref{bc1}) implies that, unlike other Rayleigh
instabilities, the rim can exchange liquid with the film and its mass
is not conserved.
\bibitem{ft_q1}
One can treat $q_1$ as another variational parameter, yet this does
not significantly improve the results.
\bibitem{ft_sin}
Identical results are obtained using a different trial function,
$\phi=\sin[k_1(u+w/2)]$ times a 3rd-order polynomial in $u$, whose
coefficients are so chosen as to satisfy the boundary conditions
(\ref{bc1}) and (\ref{bc2}) for fixed $\kappa$.
\end{thebibliography}
\end{document}